\documentclass[runningheads]{llncs}

\usepackage{eccv}

\usepackage{eccvabbrv}

\usepackage{graphicx}
\usepackage{booktabs}

\usepackage[accsupp]{axessibility}  

\usepackage{hyperref}

\usepackage{orcidlink}

\newenvironment{itemizedot}{\begin{itemize} }{\end{itemize}}

\usepackage{amsmath}

\usepackage{latexsym}

\usepackage{multirow}

\usepackage{overpic} 

\usepackage{pict2e} 

\usepackage{tabularx}

\usepackage{xcolor}

\newcommand{\DSC}{\text{DS}^{\text{2}}\text{C}^{\text{2}}}
\newcommand{\bDSC}{\textbf{DS}^{\textbf{2}}\textbf{C}^{\textbf{2}}}

\newcommand{\tBPP}{\text{BPP}}
\newcommand{\tPete}{\text{pe}_\text{test}}
\newcommand{\tbPete}{\textbf{pe}_\textbf{test}}
\newcommand{\tPetr}{\text{pe}_\text{train}}
\newcommand{\tbPetr}{\textbf{pe}_\textbf{train}}
\newcommand{\tPSNR}{\text{PSNR}}

\begin{document}

\title{Robust End-to-End Image Transmission with Residual Learning} 


\author{Cenk M. Yetis}

\institute{Huawei Technologies, \email{cenk.yetis@huawei.com}
}

\maketitle

\begin{abstract}
Recently, deep learning (DL) based image transmission at the physical layer (PL) has become a rising trend due to its ability to significantly outperform conventional separation-based digital transmissions. However, implementing solutions at the PL requires a major shift in established standards, such as those in cellular communications. Application layer (AL) solutions present a more feasible and standards-compliant alternative. In this work, we propose a layered image transmission scheme at the AL that is robust to end-to-end (E2E) channel errors. The base layer transmits a coarse image, while the enhancement layer transmits the residual between the original and coarse images. By mapping the residual image into a latent representation that aligns with the structure of the E2E channel, our proposed solution demonstrates high robustness to E2E channel errors.
\end{abstract}

\section{Introduction}
\label{sec:Intro}

6G aims to meet extreme requirements
across all aspects of wireless communications which means we need to
move away from traditional approaches and explore new dimensions. Semantic communication (SC)
is considered a crucial step towards achieving human-like intelligence and is
expected to significantly transform all aspects of next-generation networks
{\cite{Shi2023}}. Semantics can significantly reduce the data size by discarding redundant and
non-essential information. Wireless compressed media transmission via
semantics is set to \ play a crucial role in various fields such as the
metaverse, digital twins, virtual reality, augmented reality, and the Internet
of Everything {\cite{Lan2021}}.

One of the early successes in this domain is semantic-based deep joint source
and channel coding (SB-DJSCC), which outperforms traditional separate source
and channel coding methods {\cite{Gunduz2023}}. In this line of work, semantic
representation of the data is source and channel encoded rather than the
digital bit or symbol representation of the data {\cite{Xu2023}}. Semantic
representation can be achieved in many ways including semantic segmentation,
e.g., object labeling and border lining {\cite{Lateef2019}}, latent
representation, i.e., neural network (NN) mapping {\cite{Dai2022}}, and also
many metrics can be used such as peak signal-to-noise ratio (PSNR),
multi-scale structural similarity index measure (MS-SSIM), and learned
perceptual image patch similarity (LPIPS) {\cite{Getu2023}}.

Hence, semantics-based NN solutions have recently become a key research focus
for compressed media transmission in the literature. Although some works align better with existing standards due
to their use of bitstreaming {\cite{Huang2023,Dai2023,Hu2023}}, their
solutions still require significant changes to the standards. For a more
immediate standards-compliant SC solution, a more direct approach is to implement
SC in the application layer (AL) {\cite{Lan2021}}. In this work, we propose deep semantics source channel coding
($\DSC$) embedded in the AL that
is robust to end-to-end (E2E) channel errors. The robustness is achieved by mapping source data into a
latent representation which has a similar structure with the E2E channel. 

In {\cite{Lee2019,Akbari2019,Huang2021,Dong2023}}, the authors propose NN
based image compression algorithms using residual coding. Residual coding
refers to residual image transmission $r$ as an enhancement layer alongside
the base layer. $r$ complements the coarse image $x'$ transmitted from the base
layer, i.e., ${r = x - x'}$. However, the existing NN solutions with residual coding incorporate conventional compression
techniques such as BPG and JPEG into their solutions. In contrast, the
proposed $\DSC$ in this work is
a complete NN solution, which can support future SC applications due to its
input-awareness and adaptability to system changes.

$\DSC$ differs from {\cite{Lee2019,Akbari2019,Huang2021,Dong2023}} in the following ways:
\begin{enumerate}
  \item $\DSC$ is a complete NN
  solution. NN is used for encoding the residual image, referred to as block ResNet (BResNet).
  
  \item Two consecutive GANs are implemented, referred to as recursive GAN
  (RGAN).
  
  \item NN is used to improve the summation accuracy at the receiver, referred to as
  SumNet.
  
  \item $\DSC$ is a disjoint
  design of two NN blocks: RGAN and BResNet (+ SumNet at the receiver). These
  two NN blocks are trained separately.
  
  \item It is assumed that $x'$ is received without error.
\end{enumerate}
The reasons behind these design choices are intertwined.

Items $1$, $4$ and $5$: The 1st NN block generates highly compressed image
data that can be strongly protected, ensuring it is received error-free. This
means that when there are changes in the network, only the 2nd NN block needs
to be retrained or updated. However, if an error-free assumption is not
feasible, both NN blocks need retraining. Consequently, the disjoint design
provides flexibility in different scenarios.

Items $2$ and $4$: Since the 1st and 2nd NN blocks are trained independently,
the number of RGANs in the first block can be adjusted based on channel
conditions. For instance, if the output of the BResNet (2nd NN block)
encounters harsher channel conditions, the number of RGANs should be
increased. This adjustment ensures that $r$ carries less information since
$x'$ is closer to $x$. But having more RGANs increases the computational
complexity. Therefore, if $r$ faces acceptable channel conditions, the number
of RGANs should be kept low. Hence, $r$ automatically carries more information
as an enhancement layer.

Item $3$: As detailed later, a simple summation operation at the receiver is
insufficient to handle the most significant bits (MSB) and least significant
bits (LSB) during transitions between tensor and image domains. Therefore, we incorporate SumNet to improve the summation accuracy. 

\section{System Model and Related Works}

In this section, we introduce the $\DSC$ architecture and compare it to the existing NN
architectures using residual coding. We also apply quantization
of latents since it is effective to achieve highly compressed outputs while
keeping the NN architecture relatively simple {\cite{Yang2023}}. Finally, we
interleave the quantized latents before the transmission to achieve robustness
against block errors in the channel {\cite{Beltran98}}.

\subsection{$\DSC$ architecture}

\begin{figure}[!t]
\centering

\includegraphics[scale=0.23] {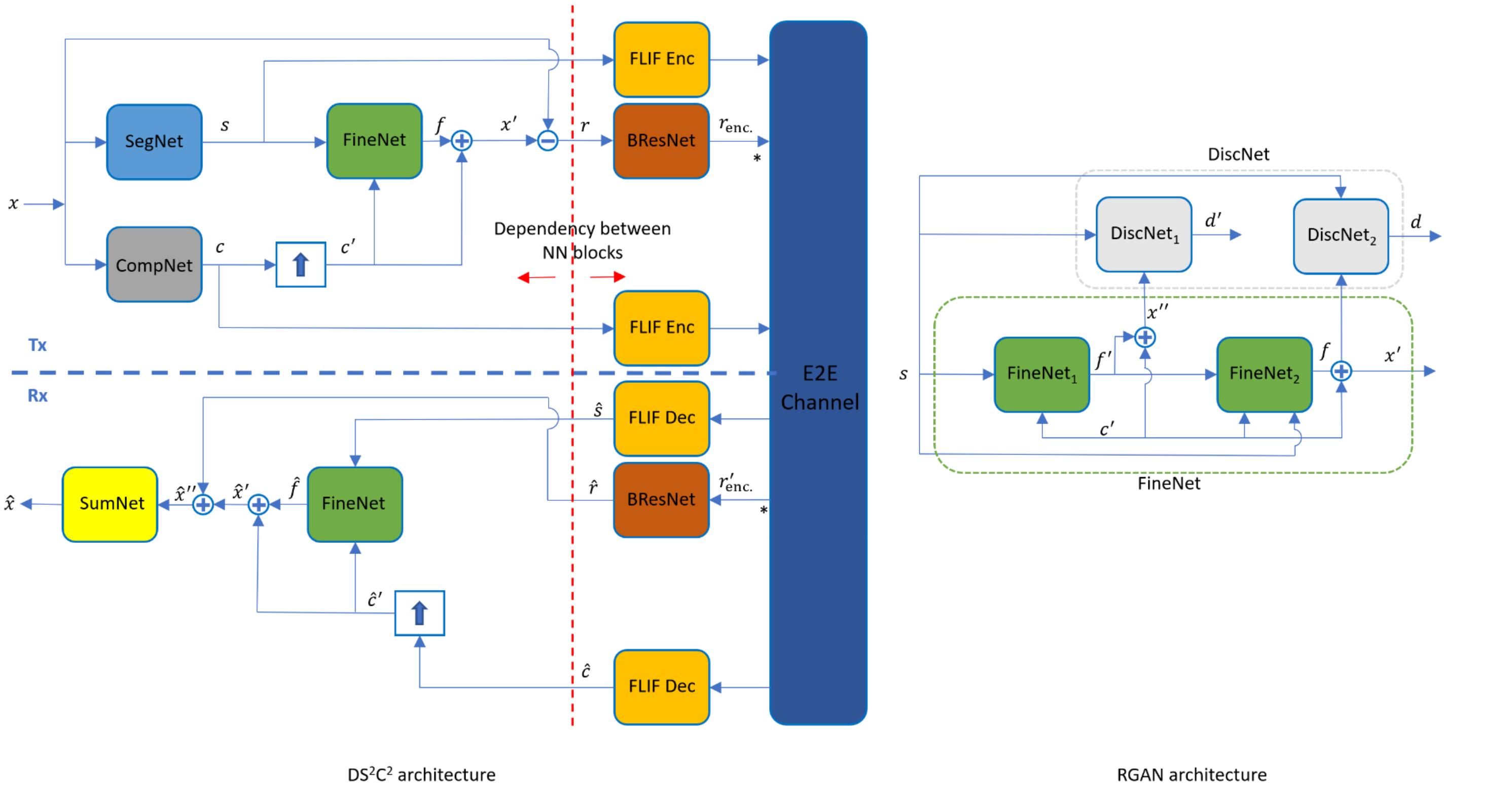}
  \caption{$\DSC$ and the inherent RGAN architectures.}
  \label{fig:DS2C2andRGANArchs}
\end{figure}

$\DSC$ architecture and the inherent RGAN architecture are shown in Fig. \ref{fig:DS2C2andRGANArchs}. In Fig. \ref{fig:DS2C2andRGANArchs}, segmentation network (SegNet) generates a semantic
image, $s$, such as labels and boundary maps. Compression network (CompNet) generates a downsampled version of the
original image, $c$, which is later upsampled to $c'$. Then $s$ and $c'$ are fed
into FineNet. The fine details in the FineNet output $f$ are summed with $c'$
to get a better estimate, $x^\prime$, coined as synthesized image. The residual image $r$ is obtained by
a subtraction operation and then encoded by block residual network (BResNet). At the
receiver, in addition to the complementary operations to the transmitter, a summation network
(SumNet) is included to improve the summation accuracy. Also, quantization and interleaving operations are applied after BResNet. This part is marked by
star in \ Fig. \ref{fig:DS2C2andRGANArchs} and further details are deferred to the next
section.

During the training phase, RGAN comprises two networks: the generator (FineNet) and the discriminator (DiscNet) networks \cite{Akbari2019}, as shown in Fig. \ref{fig:DS2C2andRGANArchs}. However, during
the test phase, DiscNet is not used. Only a ${2\text{-step}}$ RGAN is shown in
Fig. \ref{fig:DS2C2andRGANArchs}. If more steps are taken, the estimation quality of synthesized image $x^\prime$
increases at the cost of increased computational complexity.
$\text{DiscNet}_i$ is a ${3\text{-staged}}$ multi-scale discriminator network
as detailed in {\cite{Wang2018}}. The same architectures as in
{\cite{Akbari2019}} are used in SegNet, CompNet and
$\text{FineNet}_i$.

Next, we conduct a detailed benchmark comparison of $\DSC$ with existing solutions in the literature:
\begin{itemizedot}
  \item $\DSC$: RGAN to achieve
  $x'$ and $\hat{x}$. BResNet to achieve $\hat{r}$, i.e., a complete NN solution.
  
  \item {\cite{Akbari2019}}: GAN to achieve $x'$ and $\hat{x}$. BPG to achieve $\hat{r}$.
  
  \item {\cite{Lee2019}}: BPG to achieve $x'$. Residual blocks to achieve $\hat{x}$ and $\hat{r}$.
  
  \item {\cite{Dong2023}}: GAN to achieve $x'$ and $\hat{x}$. NN and JPEG2000 to achieve $\hat{r}$.
  
  \item {\cite{Huang2021}}: GAN to achieve $x'$ and $\hat{x}$. $r$ is directly sent to the channel.
\end{itemizedot}

There are three major challenges to implement a complete NN solution: 1) The
computational complexity and memory requirements are significantly increased.
2) GPU coding is notably challenging. 3) It is an open problem to be competitive in ideal channel conditions (no channel errors) against existing NN solutions without residual coding.
The first two challenges are further amplified due to the dependencies between the two NN blocks: RGAN and BResNet +
SumNet. To address this, the two NN blocks are separately optimized in this work. For SegNet, a pre-trained PSPNet proposed in {\cite{Zhao2017}} is
used. The aforementioned challenges also make it difficult to include SegNet in
the 1st NN block for joint training, i.e., RGAN + SegNet. Additionally, a
disjoint design brings the advantage of flexibility, as noted earlier.  In this work, we do not address the last challenge but we rather focus on channel error robust image transmission solution. We even prefer a fixed compression rate (i.e., bit rate) approach which allows nearly accurate predetermination of bits per pixel (BPP) before both training and inference phases. By fixing BPP, channel effects can be more certainly analyzed. However, fixed rate approach brings a major disadvantage against conventional codecs, e.g., BPG, and existing NN solutions with rate-distortion optimization. For instance, BPG is based on the intra-frame encoding of HEVC standard. Hence, BPG can significantly compress similar neighboring pixels such as flat surfaces, walls, and skies. On the other hand, $\DSC$ does not have this adaptive compression approach and fixes all images at the same BPP.  Hence, $\DSC$ falls behind in all performance metrics in ideal channel conditions but significantly outperforms in non-ideal channel conditions.

A complete NN solution can be preferable to NN solutions that integrate conventional image codecs. The preset compression rate of a conventional codec can restrict the adaptive capabilities of NN solutions. For instance, if a conventional codec is set to a high bit rate, it may impede the NN's ability to achieve the desired lower bit rate. Therefore, when targeting specific BPPs, it is crucial to select the bit rate of conventional codec with caution. Additionally, the integration of conventional codecs can limit the full potential of NN operations, such as transfer learning.

\subsection{Quantization and interleaver}

\begin{figure}[!t]
\centering
\includegraphics[scale=0.25] {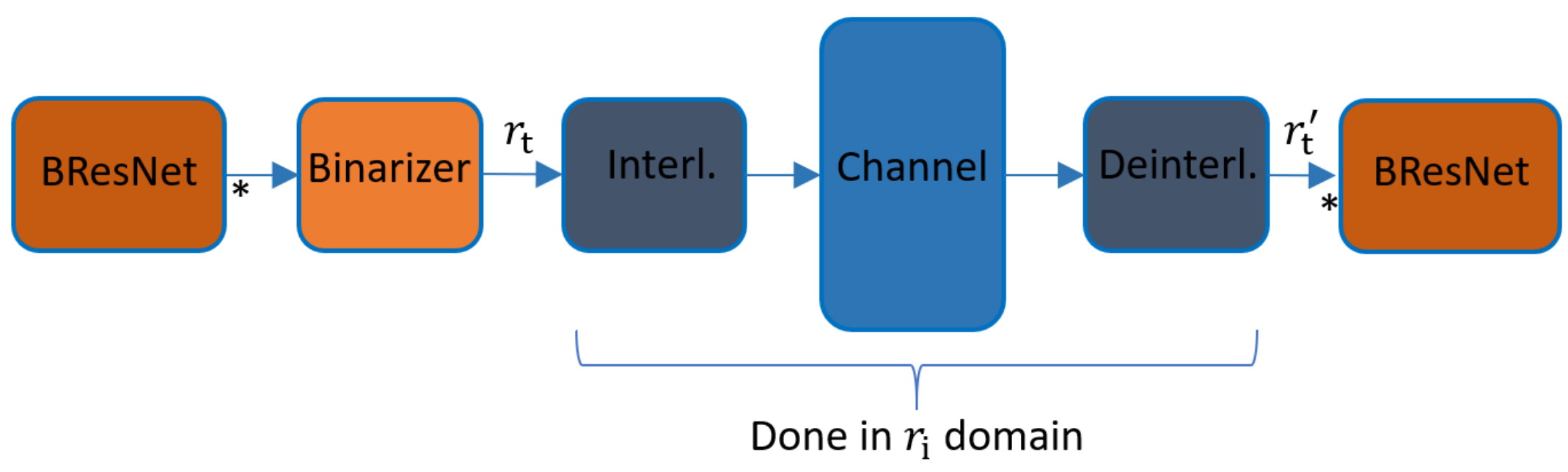}
  \caption{Binarization, interleaving, channel error modelling, and
  deinterleaving operations used in $\DSC$.}
  \label{fig:BinarizerandInterleaver}
\end{figure}

In Fig. \ref{fig:BinarizerandInterleaver}, the details quantization and interleaver in the $\DSC$ architecture are provided, expanding on the star marks in Fig.\ref{fig:DS2C2andRGANArchs}. Quantization followed by entropy coding effectively improves the rate-distortion performance. Rate-distortion optimization has been addressed
through two approaches so far: 1) Joint optimization of rate and distortion
{\cite{Cheng2020,Dai2022,Lee2019}}, and 2) Minimization of distortion given a fixed rate {\cite{Toderici2016}}. Distortion also has a more direct impact on
BPP; allowing more distortion can result in a smaller BPP. Before both training and inference phases, predetermining both the rate and BPP for the former approach and predetermining BPP for the latter approach are challenging. 

The robustness of NN against channel errors varies depending on the BPP value. As detailed later, at high BPP, channel effects are more prominent, whereas at low BPP, the BPP itself becomes the dominant factor influencing the PSNR. In this work, we use statistical
binarizer {\cite{Toderici2016}} that allows nearly accurate predetermination
of BPP before both training and inference phases. The binarizer facilitates a direct $1
/ 8$ BPP compression, and further compression is achieved by adjusting the
latent dimensions. By fixing the BPP values, we can more definitively analyze the channel effects for the proposed $\DSC$ architecture.

The binary symmetric channel (BSC) and binary erasure channel (BEC) are well-known
channel models. BSC randomly flips the bits, while the BEC has ternary outputs despite
the binary inputs. In this work, we assume BEC model with a bit
probability of flipping from $1$ to $0$ (or equivalently from $0$ to $1$). This can model dropped packets
and corrupted symbols {\cite{Farsad2018,Shao2020}}. Furthmore, we assume block BEC (BBEC) in our work because, according to 5G NR specifications, a code block can be a maximum of $1056$ bytes. To mitigate the severity of block losses or block errors, a pixel-level
interleaver is used before the transmission {\cite{Beltran98}}.

We model the BBEC with a probability representing the percentage of bits being flipped from $1$ to
$0$ during training and test phases by $\tPetr = pe_{\text{train}}\%$ and $\tPete =
pe_{\text{test}}\%$, respectively, e.g., $5\%$ of $1$s in an image are corrupted by the
channel and flipped to $0$s.

\subsection{BResNet architecture}

\begin{figure}[!t]
\centering
\includegraphics[scale=0.175] {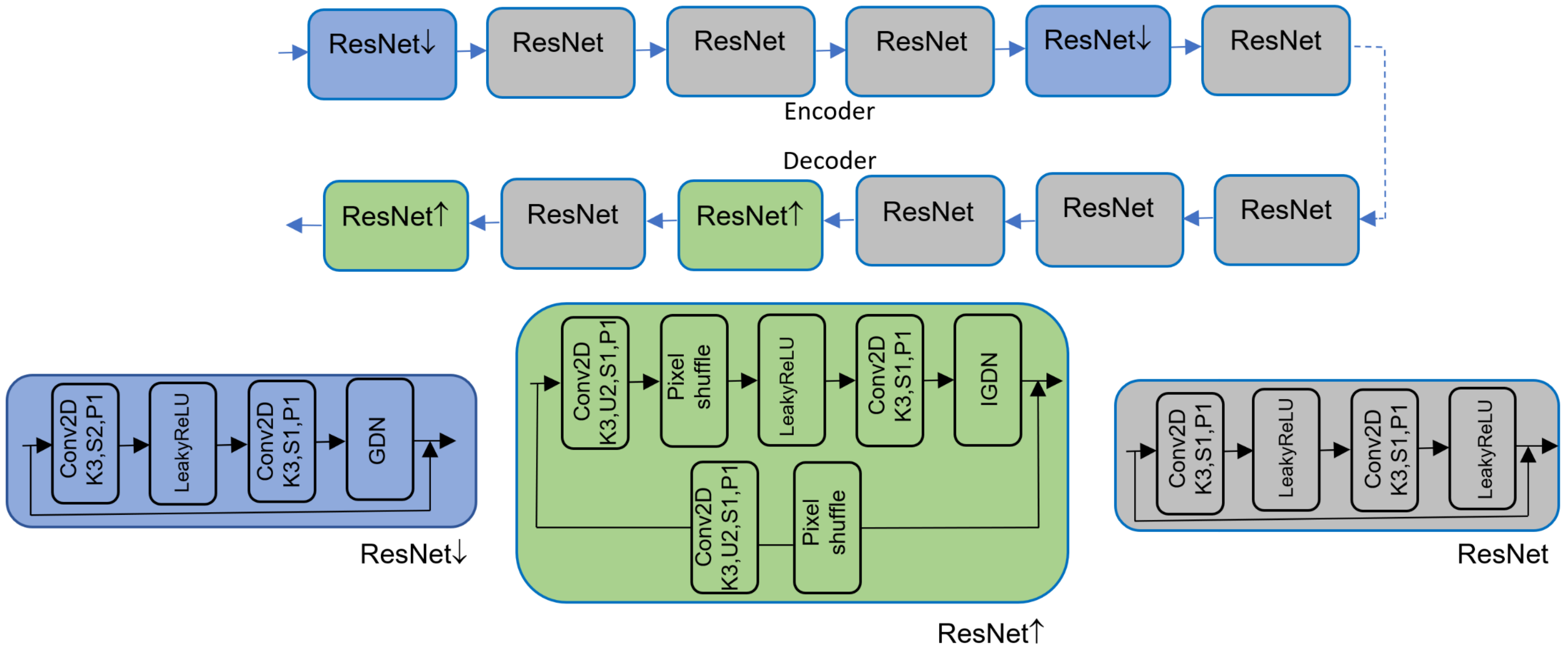}
  \caption{The details of BResNet encoder and decoder in the
  $\DSC$ architecture.}
    \label{ResNetArch}
\end{figure}

In this section, the BResNet encoder and decoder blocks used in
$\DSC$ are detailed. In Fig.
\ref{ResNetArch}, each ResNet sub-block refers to a 2D convolutional NN with a
residual connection. For the normalization of the layers, generalized
divisive normalisation (GDN) and inverse GDN (IGDN) are used. Also, a pixel
shuffle in the upsampling process is adopted instead of transpose convolution. Finaly, K,
S, P, and U refer to kernel, stride, and padding sizes, and upsampling factor, respectively. For SumNet, $3$ consecutive ResNet (gray box) sub-blocks are used.

\section{Robust Semantic Image Transmission with Residual Coding}

In this section, the proposed $\DSC$ architecture is further detailed. The loss function of RGAN is given as:
\begin{equation}
\mathcal{L}_\text{RGAN} = \mathcal{L}_\text{G} + \mathcal{L}_\text{D},
\end{equation}
where
\begin{align}
 \mathcal{L}_\text{G} & = - \sum^2_{j = 1} \sum^3_{i = 1} \log\text{D}_{j  i} (s, c', x_j') + \mathcal{L}_\text{d} \\
 \mathcal{L}_\text{D} & = - \sum^3_{i = 1} \log\text{D}_i (s, c', x) - \sum^2_{j = 1} \sum^3_{i = 1} \log (1 - \text{D}_{j i} (s, c', x_j')), \text{and}\\
 \mathcal{L}_\text{d} & = \mathcal{L}_1  +\mathcal{L}_\text{SSIM} + \mathcal{L}_\text{VGG}.
\end{align}
Here, ${\mathcal{L}_\text{d}}$ is the distance measure between the original
and estimated images, where ${\mathcal{L}_1} , \mathcal{L}_\text{SSIM},
\text{ and } \mathcal{L}_\text{VGG}$ are the well-known L1-norm, SSIM, and pre-trained VGG network loss evaluations
\cite{Akbari2019,Getu2023}, respectively.

For BResNet and SumNet, the ${\mathcal{L}_\text{d}} $ loss metric
defined above is used as well. The training of $\DSC$ is achieved in two steps. First, RGAN is trained by minimizing
$\mathcal{L}_\text{RGAN}$. Then, the pretrained RGAN is
loaded for the joint training of BResNet and SumNet by minimizing $\mathcal{L}_\text{BResNet}
+\mathcal{L}_\text{SumNet}$. In fact, we also achieved the joint training of RGAN, BResNet, and SumNet by minimizing
$\mathcal{L}_\text{RGAN} +\mathcal{L}_\text{BResNet}
+\mathcal{L}_\text{SumNet}$. However, in this case, the performance of RGAN
varies significantly across different training sessions. Specifically, for varying BPP and
channel error probability values, the PSNR between the synthesized image and the original image, $\tPSNR(x, x')$, shows considerable fluctuation. We believe this is due to the significantly enlarged NN dimensions, necessitating more training data to achieve stable results. This variability can hinder the re-training process, especially when swift updates of NNs with a small amount of data are required. Therefore, we reaffirm that disjoint optimization provides $\DSC$ with the flexibility needed in various scenarios.

It is well-known that statistical dependence in the latent representation leads to suboptimal compression performance. In \cite{Balle2018}, standard deviations of latents are predicted and used to obtain independent latents, thereby improving rate-distortion performance. However, in this work, we demonstrate that structured latent representations can enhance robustness against channel errors.

\begin{figure}[!t]
\centering
\hspace{-1.75cm}
\begin{subfigure}[t]{0.5\textwidth}
\centering
\begin{overpic}[scale=0.19] {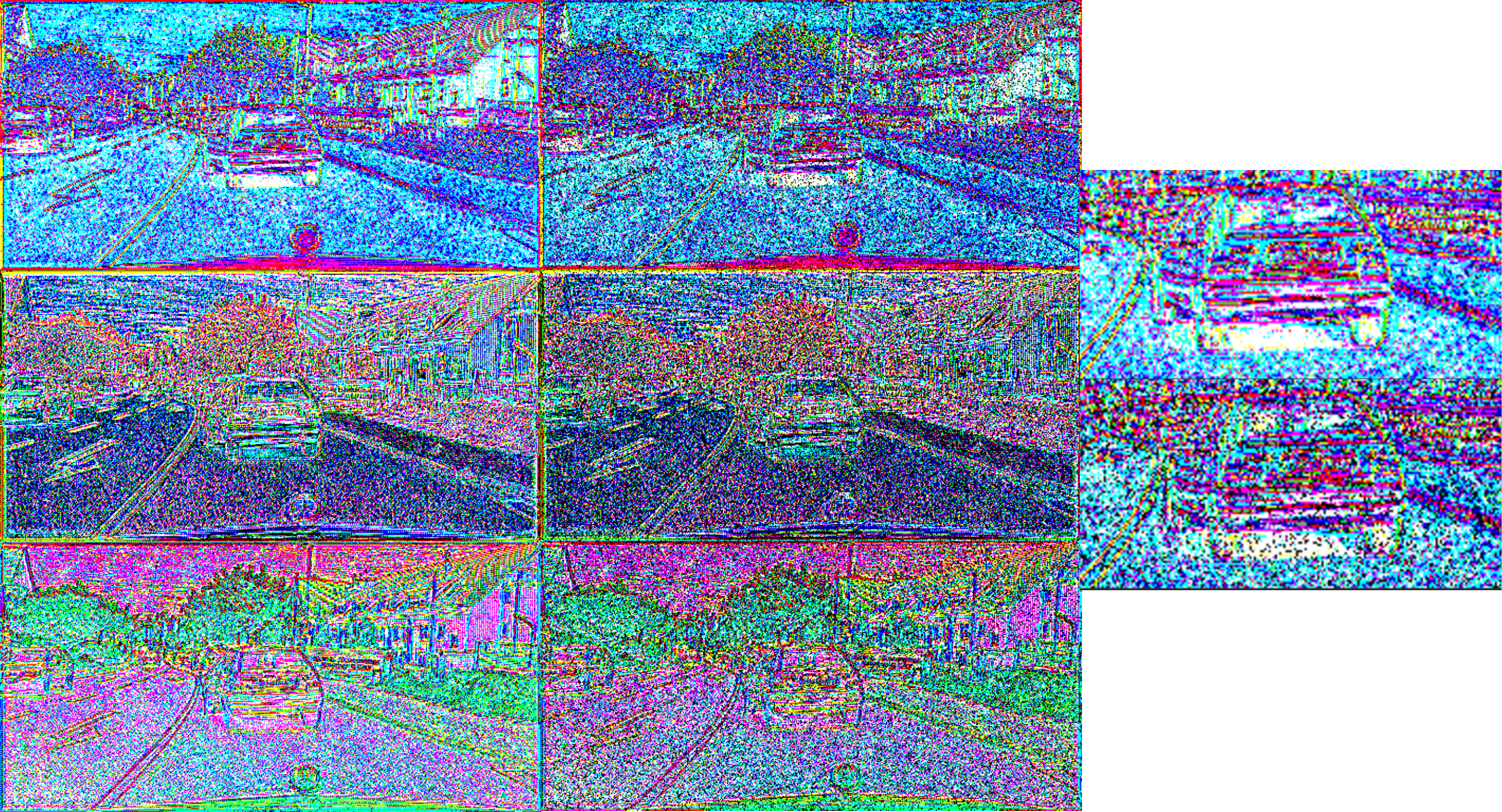}
\put(0,0){\color{yellow}\linethickness{0.2mm}\polygon(10,40)(23,40)(23,48)(10,48)}
\put(0,0){\color{yellow}\linethickness{0.2mm}\polygon(46,40)(59,40)(59,48)(46,48)}
\end{overpic}
  \caption{Cityscapes dataset.}
\end{subfigure}

\begin{subfigure}[t]{0.5\textwidth}
\centering
\begin{overpic}[scale=0.42] {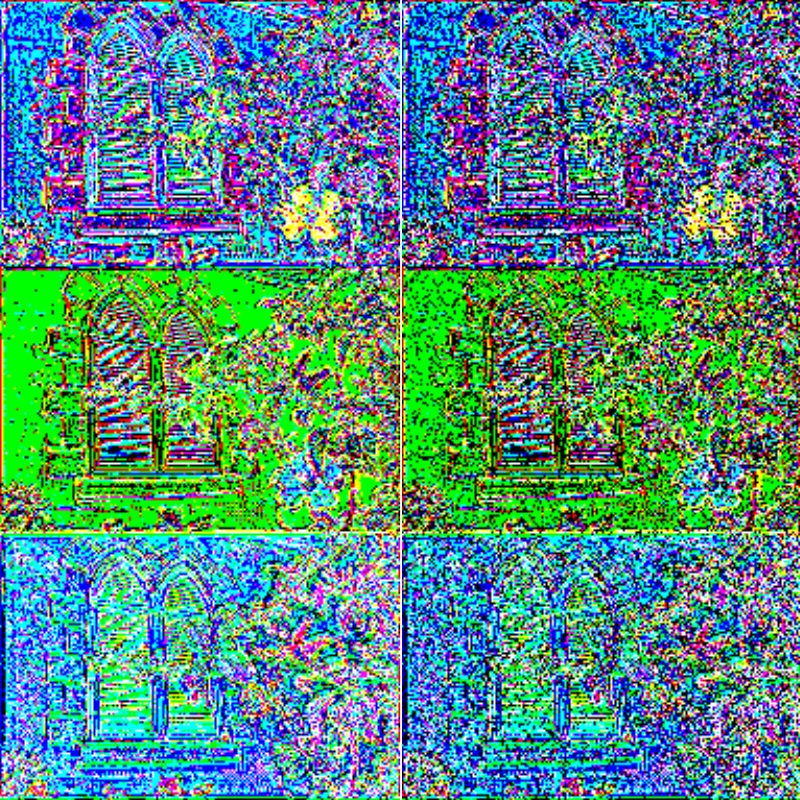}
\put(4,60){\color{black}\linethickness{0.5mm}\circle{6}}
\put(54,60){\color{black}\linethickness{0.5mm}\circle{6}}
\end{overpic}
  \caption{Kodak dataset.}
\end{subfigure}
  \caption{Similar structures observed in the latent representations of $r_{\text{i}}$ (left) and $r_{\text{i}}'$ (right) for Cityscapes and Kodak datasets.}
    \label{AllCityscapes}
\end{figure}

In Fig. \ref{AllCityscapes}, the encoded images before the interleaver
$r_{\text{i}} \text{}$ and after the deinterleaver $r_{\text{i}}'$ are shown for Cityscapes and Kodak datasets.
During the training phase, $\tPetr$ is set to $0\%$, while during the test phase, $\tPete$ is set to $16\%$. $r_{\text{i}} \text{}$ already exhibits a noisy structure.
Consequently, even after experiencing channel errors, $r_{\text{i}}'$ maintains a structure similar to $r_\text{i}$. Contrary to expectations, this indicates that
channel errors do not significantly impact $r_{\text{i}}$, as the
structure of $r_{\text{i}}$ remains largely unchanged after the E2E channel errors. In other words, the residual image is mapped into a latent representation that aligns well with the E2E channel characteristics.

\begin{figure}[!t]
\centering
\includegraphics[scale=0.5] {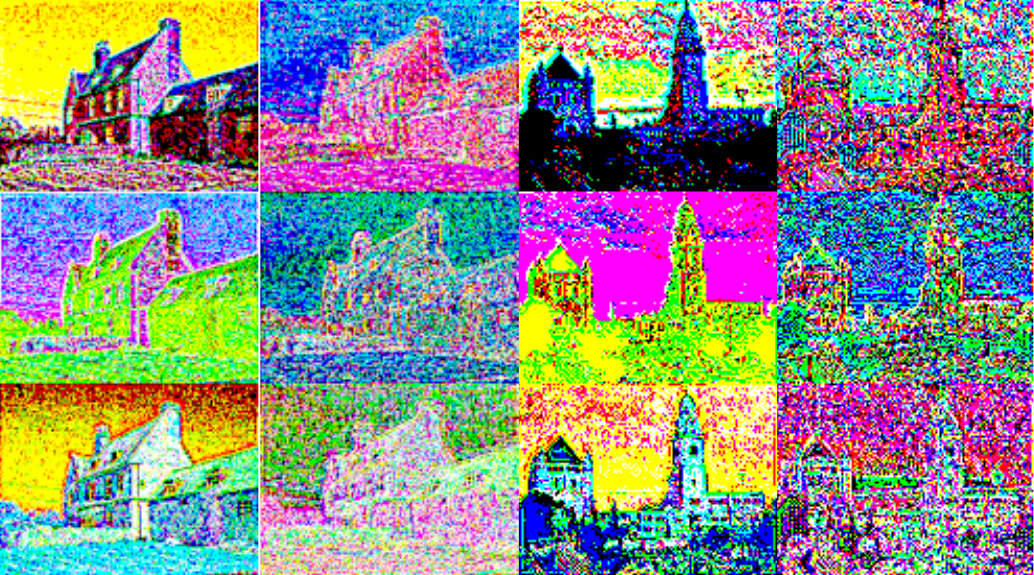}
  \caption{Structures observed in the latent representations of original image
  $x_{\text{i}}$ (left) and residual image $r_{\text{i}}$ (right) for ADE20K
  dataset.}
   \label{fig:AllADE20K}
\end{figure}

In Fig. \ref{fig:AllADE20K}, for the same pre-trained BResNet, the outputs to original image and residual image inputs are shown for the ADE20K dataset. As seen on the right side of Fig. \ref{fig:AllADE20K}, a similar structure to E2E channel is observed once again for the residual image, e.g., randomly distributed black pixels
appear in both $r_{\text{i}}$ and $r_{\text{i}}'$. However, as seen on the
left side of Fig. \ref{fig:AllADE20K}, if original image was to be encoded instead
of residual image, the encoded original image $x_{\text{i}}$ does not have a similar
structure with the E2E channel, e.g., randomly distributed black pixels do not appear.
Instead, black pixels appear more continuously (one after another). Hence,
clearly, the structures of $x_{\text{i}}$ and $x_{\text{i}}'$ are quite
different while $r_{\text{i}}$ and $r_{\text{i}}'$ have similar structures.
Therefore, encoded (compressed) residual image transmission can be more robust
against channel errors compared to encoded (compressed) original image transmission.

\section{Numerical Results}

In this section, we provide extensive numerical and visual results for
varying BPP, $\tPetr$, and $\tPete$ parameters. In addition, we benchmark
$\DSC$ with conventional compression algorithms including BPG and JPEG, and also with a competitive NN
solution in {\cite{Cheng2020}} coined here as LIC-GMM-AM. Finally, we demonstrate that
$\DSC$ can be effectively trained even with a small number of training images. 

The experiments are conducted using the Cityscapes, ADE20K, and Kodak datasets.  In the Cityscapes and ADE20K tests, $50$ images are used for each experiment, which are excluded from their respective training datasets. To demonstrate the generalization capability of $\DSC$, the pre-trained $\DSC$ model on the ADE20K dataset is applied to the classical Kodak dataset, which includes $24$ test images.  

The simulation parameters are same with \cite{Akbari2019} except that we fix the learning rate in the first $100$ epochs, and then gradually decrease the learning rate in the next $100$ epochs instead of $50$. In addition, we observe that when RGAN output $x^\prime$ in Fig. \ref{fig:DS2C2andRGANArchs} is normalized between $[-1,1]$, better overall estimation performance is achieved. Note that the original image $x$ is normalized between $[-1,1]$ instead of $[0,1]$ due to the subtraction and summation operations in $\DSC$.  Only pixel-level semantic labels $s$ obtained from SegNet are used in this work. 

As mentioned earlier, for simplicity, $\tPetr$ and $\tPete$ values set the percentage of $1$s corrupted and flipped to $0$s per image by the channel. In addition, we assume block error size as $100$ bytes instead of $1056$. Because as the compressed residual image size gets smaller, e.g., for $0.11$ BPP, randomly flipping $x\%$ of $1$s to $0$s becomes a computationally costly task. For large block sizes, strong pixel-level interleaver is needed, i.e., the interleaved flipped errors should be far from each other. In all simulations, BBEC channel model is used. BBSC channel model yields lower PSNR results compared to BBEC since the former is a more detrimental channel model. 

For the interleaving, channel error modelling, and deinterleaving operations, we do
transitions between the tensor domain $r_{\text{t}} \in \{ - 1., 1. \}$ (floats)
and image domain $r_{\text{i}} \in \{ 0, 255 \}$. The deinterleaver output is in 
the tensor domain $r'_{\text{t}}$ as  as seen in Fig. \ref{fig:BinarizerandInterleaver}. Hence, E2E NN training only sees
the transition from $r_{\text{t}}$ to $r'_{\text{t}}$ to achieve learning.
Since binarizer is used, there are only two outputs. Therefore, the operations
can be directly done in tensor domain, but for the completeness of
simulations, they are done in the image domain.

Under ideal channel conditions at $0.3$ BPP, incorporating RGAN into the architecture increases PSNR by approximately $1.5$ dB. Adding SumNet further increases PSNR by an additional  $1.5$ dB. Therefore, integrating RGAN and SumNet results in a total PSNR improvement of around $3$ dB. Without an interleaver, block errors of 100 bytes can significantly degrade the PSNR. The absence of an interleaver causes a loss of structural similarity between the encoded image and channel errors. Specifically, while the encoded image contains scattered black pixels, the channel errors result in continuous black regions because the interleaver is removed. Additionally, convolutional neural networks (CNNs) become less effective as they frequently receive inputs with all black pixels in local receptive fields, leading to repeated exposure to all black pixel inputs.

For completeness of analysis, in Section \ref{sec:VaryingBPP}, the results of $\DSC$ with joint design of two NN blocks are presented. Then these results are used to benchmark $\DSC$ with the conventional and NN codecs in Sections \ref{sec:BenchConv} and \ref{sec:BenchNN}.  In Section \ref{sec:LowNumber}, the results of $\DSC$ with disjoint design of two NN blocks are presented.

\subsection{Varying BPP, $\tPetr$, and $\tPete$ parameters} \label{sec:VaryingBPP}

In this section, numerical results for varying BPP, $\tPetr$, and $\tPete$
parameters are presented. Since stochastic binarizer is used, $1 / 8$ compression is
automatically achieved. To vary BPP, latent dimensions are varied.

In Table \ref{tab:VaryingCityscapes}, results for Cityscapes dataset are
shown. In the table, $\Delta$ reads the PSNR difference between $\tPete\!\! = 0$ and $\tPete\!\! = 16$ for the
same $\tPetr$ value. As seen in the results, without channel error training (i.e., ${\tPetr = 0}$), the loss may reach as high as $4$
dB. However, with channel error training (i.e., ${\tPetr > 0}$), the loss can be reduced to as little as $0.2$ dB.
The significance of training with channel errors diminishes as BPP decreases. Lower
BPP values result in compromised encoding and decoding quality. Consequently,
when a pixel value is inaccurately represented due to encoding and decoding at low
BPP, correcting channel errors proves ineffective. In this work, tables are preferred over plots for presenting numerical results. Given the page limitations, plots cannot be displayed at a sufficient size, making it difficult to accurately interpret the gains and losses.

As seen in the last column, PSNR of synthesized image varies from $27.76$ to $31.01$.  We believe the reason for this variance is due to the significantly enlarged NN dimensions for the joint design of two NN blocks. Hence, disjoint design brings flexibility, and also convenience to choose the best performing RGAN as a BResNet input. In this case,  $\tPetr\!\! = 16$ experiment achieves the highest ${\textrm{PSNR}=31.01}$ for RGAN, and it can be used as a pre-trained input to BResNet to achieve higher estimated image PSNR. These results are presented in Section \ref{sec:LowNumber} for the Cityscapes dataset.

\begin{table}[!t]
  \begin{center}
    \caption{PSNR results for varying BPP, $\tPetr$, and $\tPete$ parameters for Cityscapes
  dataset.}
  \begin{tabular}{|c c c c c c|}\hline
$\tbPetr$&\textbf{BPP}&$\mathbf{\tbPete\!\!=0}$&$\mathbf{\tbPete\!\!=16}$ & $\boldsymbol{\Delta}$ & $\mathbf{PSNR(\boldsymbol{x,x^\prime})}$\\ \hline
0 & \multirow{2}{*}{$0.11$} & $34.68$ & $33.03$ & $1.65$ & $30.82$ \\
8 &                                          & $34.16$ & $33.94$ & $\color{green}\mathbf{0.22}$ & $29.95$ \\ \hline

0 & \multirow{3}{*}{$0.24$} & $36.58$ & $33.65$ & $2.93$ & $28.8$ \\
8 &                                          & $36.64$ & $36.00$ & $0.64$ & $29.12$ \\
16 &                                        & $37.04$ & $36.56$ & $0.48$ & $\mathbf{31.01}$ \\ \hline

0 & \multirow{3}{*}{$0.54$} & $39.49$ & $35.44$ & $\color{red}\mathbf{4.05}$ & $27.76$ \\
8 &                                          & $39.6$    & $38.54$ & $1.06$ & $28.7$ \\
16 &                                        & $39.62$ & $38.88$ & $0.74$ & $30.37$ \\ \hline
  \end{tabular}
      \label{tab:VaryingCityscapes}
 \end{center}
\end{table}

The results for ADE20K and Kodak datasets are shown in Table \ref{tab:VaryingADE20KandKodak}. PSNRs of RGAN outputs are observed to vary lesser for ADE20K and Kodak datasets compared to Cityscapes dataset. The former datasets contain more details, e.g., close-up human heads,  compared to the latter. Since it is more difficult to estimate the details, RGAN converges to one of the many closely spaced local optimas. 

\begin{table}[!t]
  \begin{center}
    \caption{PSNR results for varying BPP, $\tPetr$, and $\tPete$ parameters for ADE20K and Kodak
  datasets.}
  \begin{tabular}{|c | c c c c c c|}\hline
\textbf{ADE20K} & $\tbPetr$&\textbf{BPP}&$\mathbf{\tbPete\!\!=0}$&$\mathbf{\tbPete\!\!=16}$ & $\boldsymbol{\Delta}$ & $\mathbf{PSNR(\boldsymbol{x,x^\prime})}$\\ \hline
& 0 & \multirow{2}{*}{$0.14$} & $25.79$ & $24.50$ & $1.29$ & $22.05$ \\
& 8 &                                          & $25.69$ & $25.47$ & $\color{green}\mathbf{0.22}$ & $21.91$ \\ \cline{2-7} 

& 0 & \multirow{2}{*}{$0.26$} & $28.99$ & $25.91$ & $\color{red}\mathbf{3.08}$ & $22.03$ \\
& 8 &                                          & $28.81$ & $28.18$ & $0.63$ & $22.08$ \\ \hline \hline
\textbf{Kodak} & $\tbPetr$&\textbf{BPP}&$\mathbf{\tbPete\!\!=0}$&$\mathbf{\tbPete\!\!=16}$ & $\boldsymbol{\Delta}$ & $\mathbf{PSNR(\boldsymbol{x,x^\prime})}$\\ \hline
& 0 & \multirow{2}{*}{$0.14$} & $25.3$ & $23.67$ & $1.63$ & $20.95$ \\
& 8 &                                          & $24.67$ & $24.5$ & $\color{green}\mathbf{0.17}$ & $21.11$ \\ \cline{2-7} 

& 0 & \multirow{2}{*}{$0.26$} & $28.25$ & $25.03$ & $\color{red}\mathbf{3.22}$ & $21.04$ \\
& 8 &                                          & $27.95$ & $27.32$ & $0.63$ & $21.26$ \\ \hline
  \end{tabular}
      \label{tab:VaryingADE20KandKodak}
 \end{center}
\end{table}

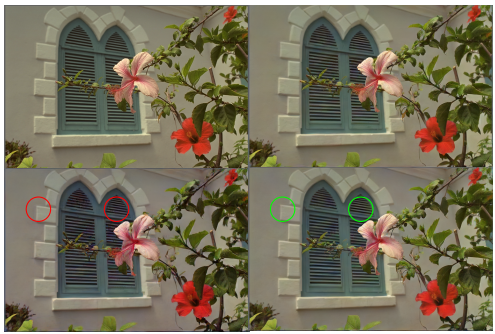
\begin{figure}[!t]
\centering
\begin{overpic}[scale=0.12] {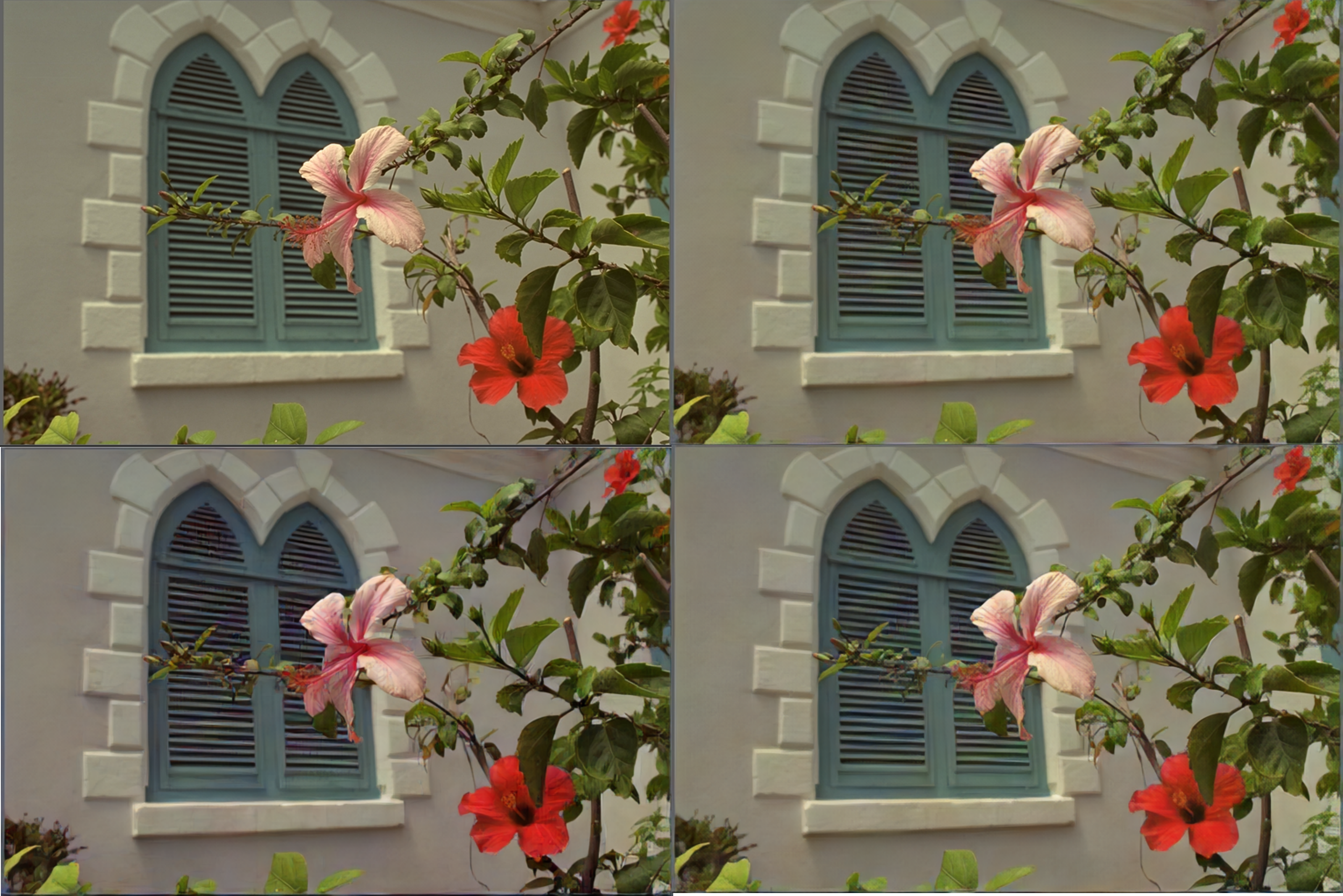}
\put(7,25){\color{red}\circle{5}}
\put(23,25){\color{red}\circle{5}}
\put(57,25){\color{green}\circle{5}}
\put(73,25){\color{green}\circle{5}}
\end{overpic}
  \caption{Original and estimated images of $\DSC$ for Kodak dataset at $0.26~ \tBPP$. Left-top: Original image. Right-top: ${\tPetr=0,\tPete=0}$. Left-bottom: ${\tPetr=0,\tPete=16}$. Right-bottom: ${\tPetr=8,\tPete=16}$.}
   \label{fig:AllKodakFlower}
\end{figure}

Next, the benefits of channel error training are visually shown in Fig. \ref{fig:AllKodakFlower}. Notice the overall hue change, i.e., more blue hue, in the  left-bottom image and the correct hue in the right-bottom image. Also some estimation accuracies are better, such as the shutters, when $\DSC$ is trained against channel errors.

\begin{figure}[!t]
\centering
\begin{subfigure}[t]{0.5\textwidth}
\centering
\includegraphics[scale=0.11] {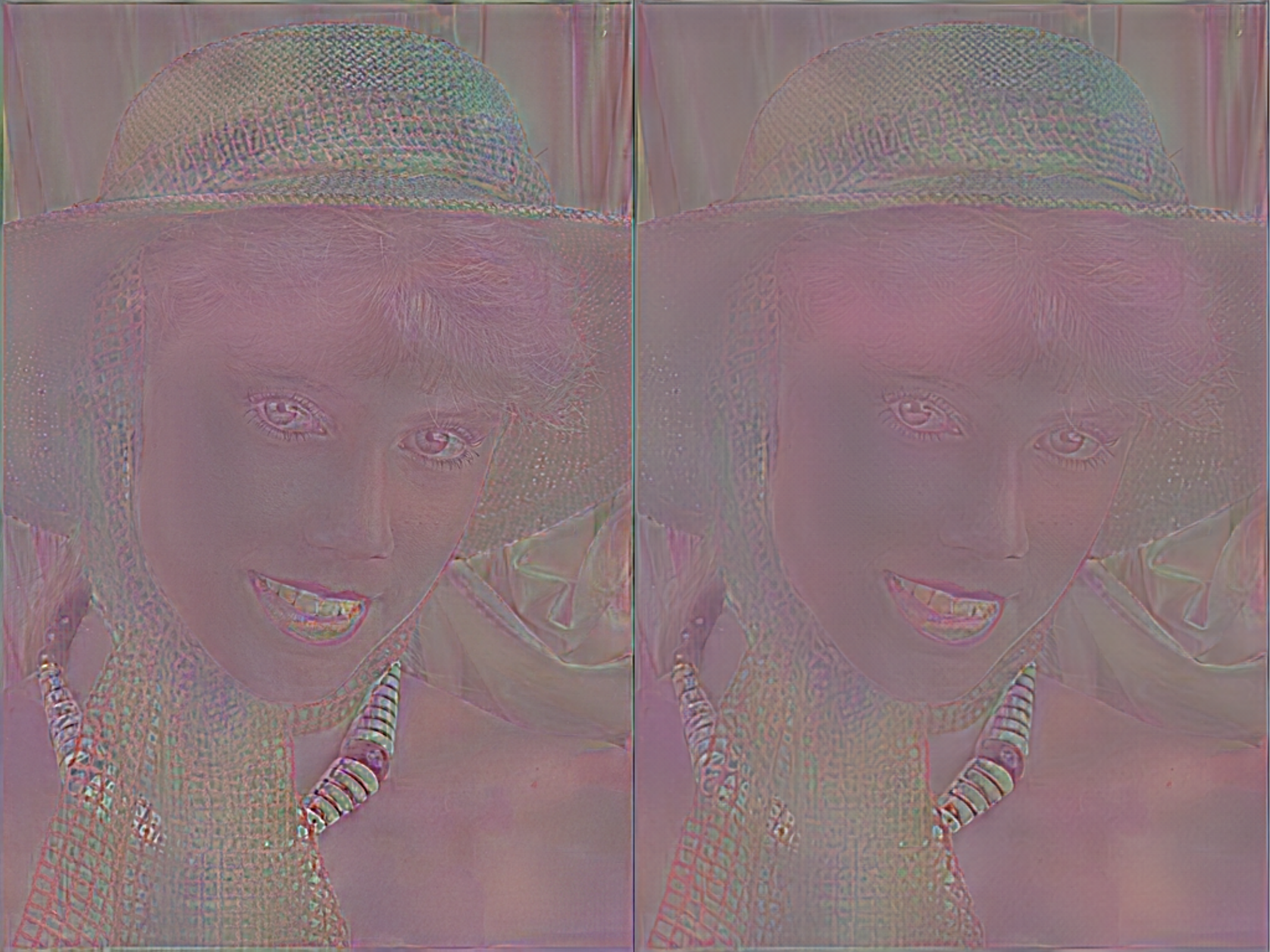}
  \label{fig:All_kodim04_residual_image}
\end{subfigure}

\begin{subfigure}[t]{0.5\textwidth}
\centering
\includegraphics[scale=0.11] {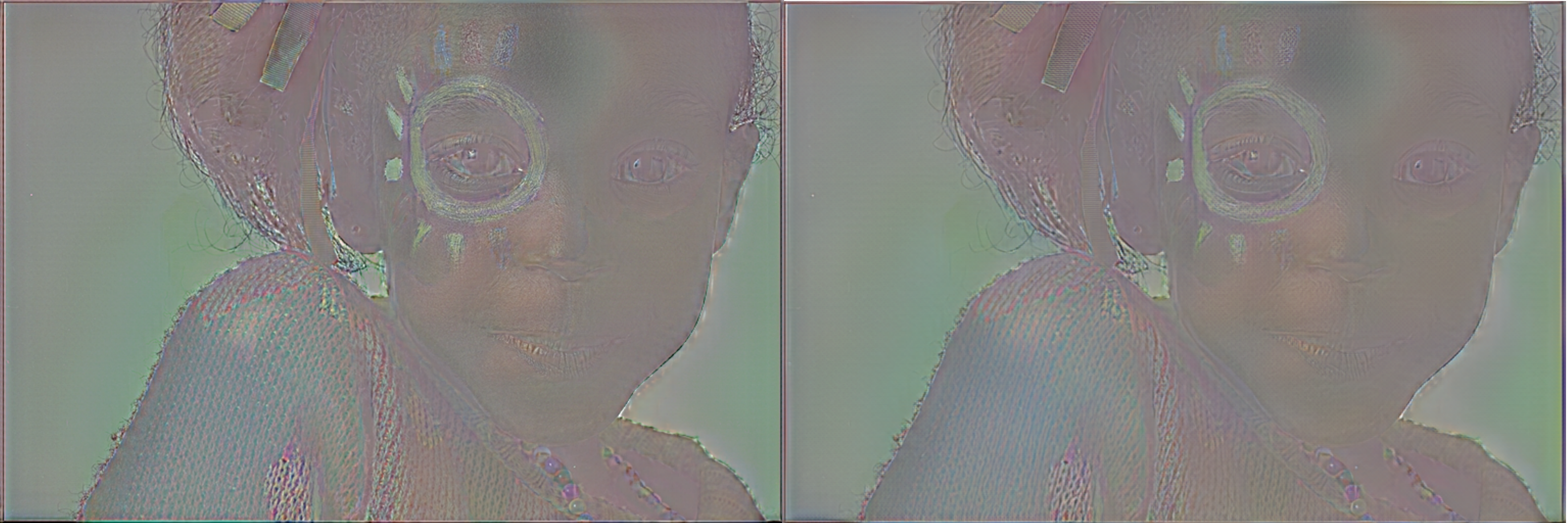}
    \label{fig:All_kodim15_residual_image}
\end{subfigure}
  \caption{Original (left) and estimated (right) residual images for Kodak dataset for $\tPetr=0$ and $\tPete=0$ at $0.26~ \tBPP$. }
      \label{fig:All_kodim_residual_image}
\end{figure}

Next, the visual results for residual image estimation are presented in Fig. \ref{fig:All_kodim_residual_image}. The proposed BResNet effectively estimates the residual image and captures important details that the RGAN output misses. 

\subsection{Benchmarking to conventional algorithms} \label{sec:BenchConv}
In this section, $\DSC$ is benchmarked to conventional algorithms, BPG and JPEG, in ideal channel conditions.  The benchmarks for Cityscapes, ADE20K and Kodak datasets are shown in Table \ref{tab:ConvenDatasets}. The results of $\DSC$ under ideal channel conditions (i.e., ${\tPete=0}$ and ${\tPetr=0}$) are identical to those presented in Tables \ref{tab:VaryingCityscapes} and \ref{tab:VaryingADE20KandKodak}.

As seen in Table \ref{tab:ConvenDatasets}, for BPPs lower than $0.54$, $\DSC$ achieves higher PSNR than JPEG, and at low BPP, $\DSC$ achieves significantly higher PSNR than JPEG.  BPG has a significant PSNR advantage over $\DSC$ at every BPP. The reason is that BPG is based on the intra-frame encoding of HEVC standard. Hence, BPG can significantly compress similar neighboring pixels such as flat surfaces, walls, and skies. On the other hand, $\DSC$ does not have this adaptive compression approach and fixes all images at the same BPP. Nevertheless, both BPG and JPEG are prune to channel errors and they have zero capabilities to correct any E2E channel errors.

\begin{table}[!t]
  \begin{center}
    \caption{BPG and JPEG benchmarks for Cityscapes, ADE20K, and Kodak datasets in ideal channel conditions.}
  \begin{tabular}{| c |c c | c c | c c|}\hline
\textbf{Cityscapes} & \textbf{BPP} & \textbf{BPG} & \textbf{BPP} & $\bDSC$ & \textbf{BPP} & \textbf{JPEG} \\ \hline
& $0.12$ & $40.51$ & $0.11$ & $34.68$ & $0.11$ & 25.48 \\ 
& $0.22$ & $43.12$ & $0.24$ & $36.58$ & $0.24$ & 35.99 \\ 
& $0.54$ & $46.86$ & $0.54$ & $39.49$ & $0.54$ & 39.8 \\ \hline \hline
\textbf{ADE20K} & \textbf{BPP} & \textbf{BPG} & \textbf{BPP} & $\bDSC$ & \textbf{BPP} & \textbf{JPEG} \\ \hline
& $0.14$ & $32.31$ & $0.14$ & $25.79$ & $0.16$ &21.11 \\ 
& $0.26$ & $36.90$ & $0.26$ & $28.99$ & $0.26$ & 25.10 \\ \hline \hline
\textbf{Kodak} & \textbf{BPP} & \textbf{BPG} & \textbf{BPP} & $\bDSC$ & \textbf{BPP} & \textbf{JPEG} \\ \hline
& $0.15$ & $29.84$ & $0.14$ & $25.3$ & $0.14$ &21.17 \\ 
& $0.25$ & $31.63$ & $0.26$ & $28.25$ & $0.26$ & 25.97 \\ \hline
  \end{tabular}
      \label{tab:ConvenDatasets}
 \end{center}
\end{table}

\subsection{Benchmarking to existing NN architecture} \label{sec:BenchNN}

\begin{figure}[!t]
\centering
\begin{overpic}[scale=0.09] {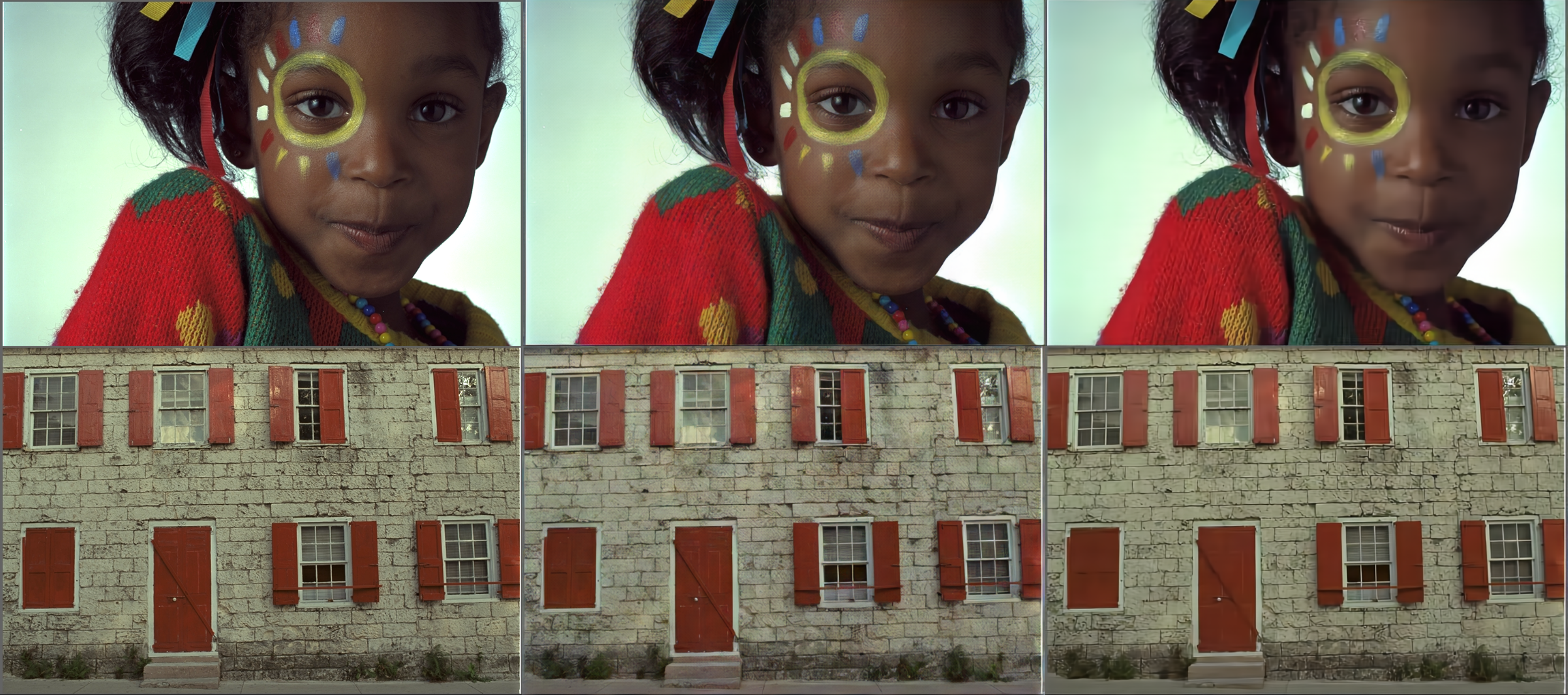}
\put(58,42){\color{green}\circle{5}}
\put(53.5,39.5){\color{green}\circle{3.55}}
\put(56,29){\color{green}\circle{3}}

\put(91,42){\color{red}\circle{5}}
\put(86.5,39.5){\color{red}\circle{4}}
\put(89,29){\color{red}\circle{3}}

\put(36.4,8){\color{green}\circle{4}}
\put(45,7){\color{green}\oval(4,8)}

\put(69.7,8){\color{red}\circle{4}}
\put(78.3,7){\color{red}\oval(4,8)}

\end{overpic}
  \caption{Original (left) and estimated images of $\DSC$ (middle) and LIC-GMM-AM (right) for Kodak dataset at $0.26~ \tBPP$. }
   \label{fig:LICGMMAMBenchmark}
\end{figure}
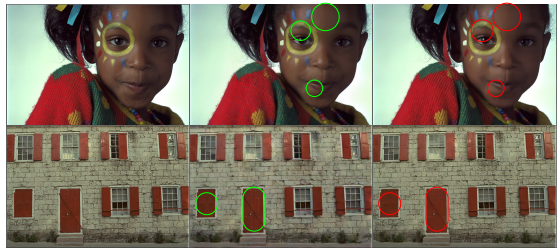

In this section, $\DSC$ is benchmarked to a competitive NN solution in {\cite{Cheng2020}} coined here as LIC-GMM-AM. At $0.26~ \tBPP$, for ${\tPetr =0}$ and ${\tPete =0}$, $\DSC$ and  LIC-GMM-AM achieve $28.25$ and $30.61$ PSNR for the Kodak dataset, respectively. In terms of MS-SSIM,  $\DSC$ and  LIC-GMM-AM achieve $0.9526$ (equivalently $13.24$ dB, ${-10\log_{10}(1-<\text{scalar value}>)}$) and $0.9605$ (equivalently $14.02$ dB), respectively. For LPIPS, $\DSC$ and LIC-GMM-AM achieve $0.0176$ and $0.0108$, respectively. Although $\DSC$ falls behind LIC-GMM-AM in these metrics, as seen in Fig. \ref{fig:LICGMMAMBenchmark}, LIC-GMM-AM has a smoothing (softening) defect on estimated images which causes the details to be lost. In contrast, $\DSC$ can pay attention to details due to the residual image, which can capture the details missed by the RGAN. But while the details are captured well, checkerboard artifacts are also created in the images. To eliminate this artifact, we believe BResNet needs to be designed more carefully. Hence, with more carefull BResNet and SumNet designs, and with rate-distortion optimization instead of fixed rate, we believe $\DSC$ can perform better than LIC-GMM-AM in ideal channel conditions. However, as previously noted, this work focuses on robust image transmission under channel errors, leaving ideal channel conditions for future research. Nevertheless, similar to conventional compression algorithms,  LIC-GMM-AM has also no channel error correction capability unless it is trained against channel errors. For ${\tPetr=0}$ and ${\tPete=8}$, PSNR of LIC-GMM-AM drops already to $4.92$ from $30.61$ at $0.26~ \tBPP$, and it achieves a negative PSNR value of $-1.18$ for ${\tPete=16}$. When channel errors are introduced during training, for ${\tPetr=8}$ and ${\tPete=16}$, it can achieve only $24.14$ PSNR. On the other hand, $\DSC$ achieves PSNR values of $25.03$ and $27.32$ for ${\tPetr=0}$, ${\tPete=16}$ and ${\tPetr=8}$, ${\tPete=16}$, respectively. The results of $\DSC$ for KODAK dataset are identical to those presented in Table \ref{tab:VaryingADE20KandKodak}.

In other words, if $\DSC$ is trained in the absence
of channel errors, but channel errors are experienced during the inference
phase, only $11.4\%$ loss in PSNR is observed. On the other hand, the PSNR loss of LIC-GMM-AM is
observed to be $83.9\%$ and $104\%$ at $\tPete=8$ and $\tPete=16$, respectively. Although $\DSC$ is not trained to handle channel errors, the
similarity between the learned structures during mapping and de-mapping and
the E2E channel prevents any significant loss in PSNR. When channel errors are
introduced during training, the loss in PSNR decreases to $3.3\%$ for
$\DSC$ and to $21.14\%$ for LIC-GMM-AM. The results for the loss percentages are summarized in Table \ref{tab:DSCvsLICGMMAM}.

\begin{table}[!t]
  \begin{center}
    \caption{$\DSC$ vs. LIC-GMM-AM for KODAK dataset at $0.26$ BPP. Benchmark case: $\tPetr\!\! = 0$, $\tPete\!\! = 0$ (no channel error is introduced during both training and testing phases). Trained without channel errors: $\tPetr\!\! = 0$, $\tPete\!\! = 16$ (no channel error is introduced during training phase). Trained with channel errors: $\tPetr\!\! = 8$, $\tPete\!\! = 16$ (channel error is introduced during both training and testing phases).}
  \begin{tabular}{| c | c | c | c |}\hline
\multirow{2}{*}{\textbf{PSNR (Loss} $\boldsymbol{\%}$\textbf{)}} & \multirow{2}{*}{\textbf{Benchmark}} &\textbf{Trained} & \textbf{Trained}  \\ 
                                                                                       &  & \textbf{ without channel errors} & \textbf{with channel errors}  \\ \hline
$\bDSC$ &28.25  & 25.03 $\color{green}\mathbf{(11.4\%)}$ & 27.32 $\color{green}\mathbf{(3.3\%)}$  \\ \hline
\textbf{LIC-GMM-AM} & 30.61 & -1.18 $\color{red}\mathbf{(104\%)}$ & 24.14 $\color{red}\mathbf{(21.14\%)}$  \\  \hline
  \end{tabular}
      \label{tab:DSCvsLICGMMAM}
 \end{center}
\end{table}

Training with channel errors for LIC-GMM-AM is not as effective as $\DSC$ because the encoded image structure differs significantly from the channel error structure for LIC-GMM-AM. To our best knowledge, there is no existing NN architecture with a structured encoded image since structure reduces the rate-distortion performance  \cite{Balle2018} as mentioned earlier. Hence, similar to  LIC-GMM-AM, they also perform poorly against channel errors. The solution in \cite{Akbari2019} has an additional drawback, as the residual image is BPG encoded, and conventional codecs, as mentioned earlier, lack any capability to correct E2E channel errors. $\DSC$ builds upon the foundation of {\cite{Akbari2019}}, incorporating several innovative modifications. BPG-encoded residual coding is replaced with BResNet, while the latents are quantized and interleaved, and  channel errors are simulated during both the training and testing phases. Furthermore, RGAN is introduced as a replacement for GAN, enhancing the model's performance.

Key challenges in image processing can be addressed through joint solutions via E2E training {\cite{Xing2021}}. The promising results of this work open avenues for further research to derive new design principles for jointly correcting protocol layer effects at AL. For instance, exploring whether other PL effects, such as IQ imbalance and frequency offset, in addition to channel errors, can also be jointly mitigated through E2E training at the AL.

\subsection{Low number of training images}\label{sec:LowNumber}

Swift update of NNs is a critical issue when an immediate update is needed due to changing conditions or insufficient training for particular conditions. For re-training, updating NNs with much less NN inputs compared to the initial training is very important for a quick response to avoid losing quality of experience (QoE). Low number of NN inputs for updating is also advantageous when collecting new training inputs which can be challenging and time consuming. 

\begin{figure}[!t]
\centering
\begin{subfigure}[t]{0.5\textwidth}
\centering
\includegraphics[scale=0.04] {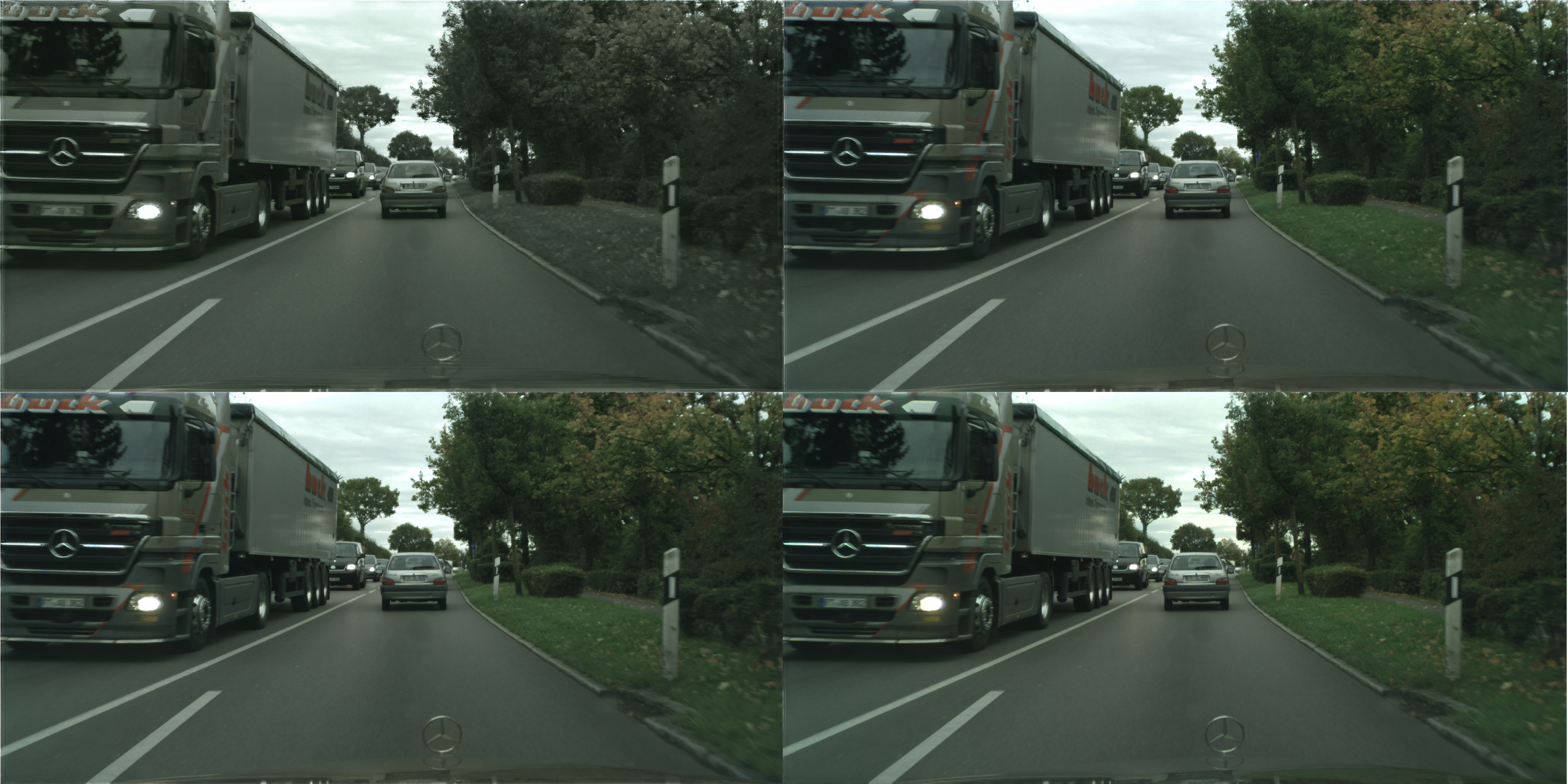}
  \caption{Only vehicles.}
  \label{fig:All01}
\end{subfigure}

\begin{subfigure}[t]{0.5\textwidth}
\centering
\includegraphics[scale=0.04] {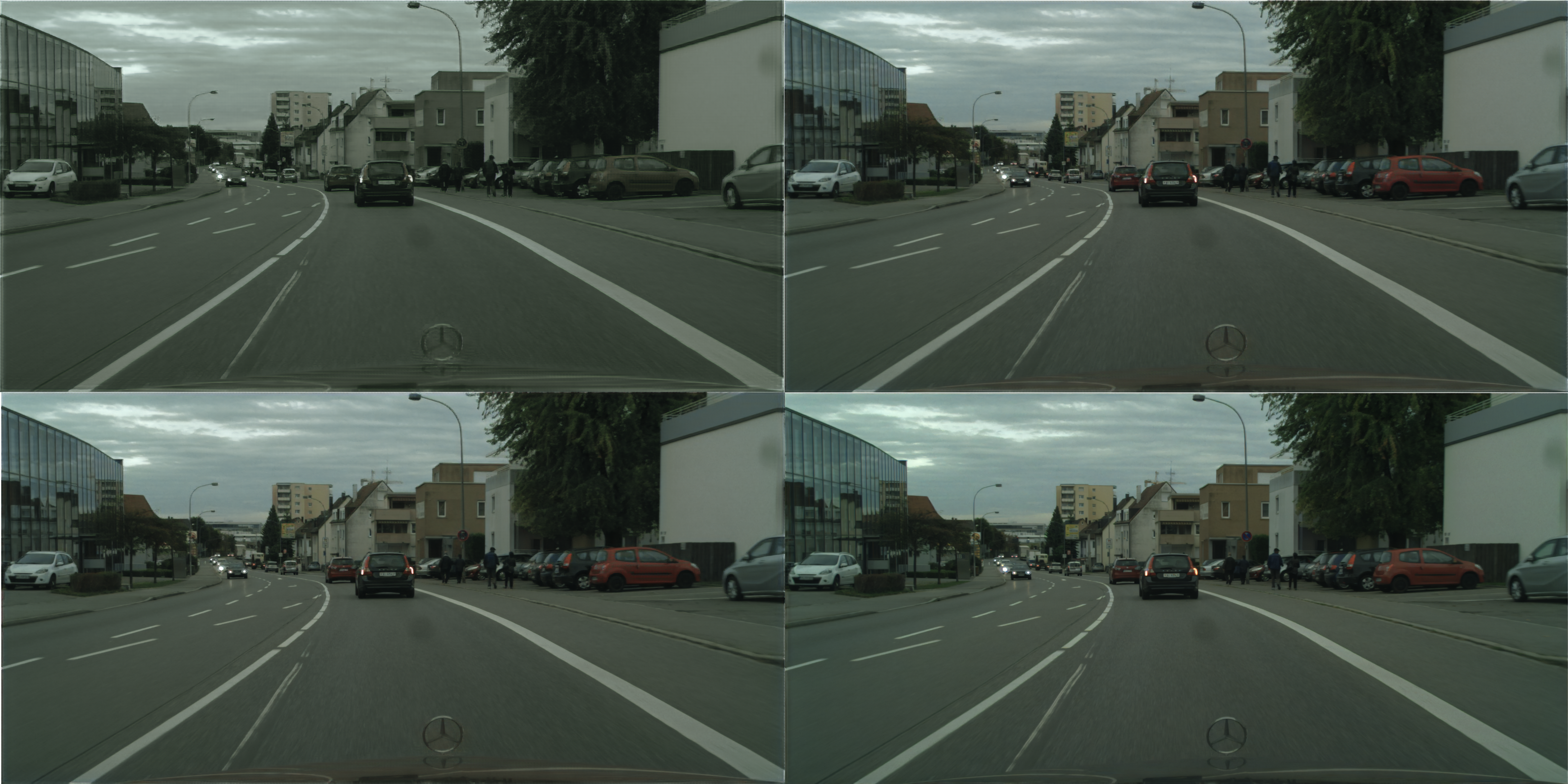}
  \caption{Vehicles and pedestrians.}
    \label{fig:All05}
\end{subfigure}
  \caption{Estimated images of $\DSC$ for Cityscapes dataset at $0.54~ \tBPP$ for varying number of training images. Left-top: $16$. Right-top: $88$. Left-bottom: $176$. Right-bottom: $2968$.}
  \label{fig:All01and05}
\end{figure}

\begin{figure}[!t]
\centering
\includegraphics[scale=0.15] {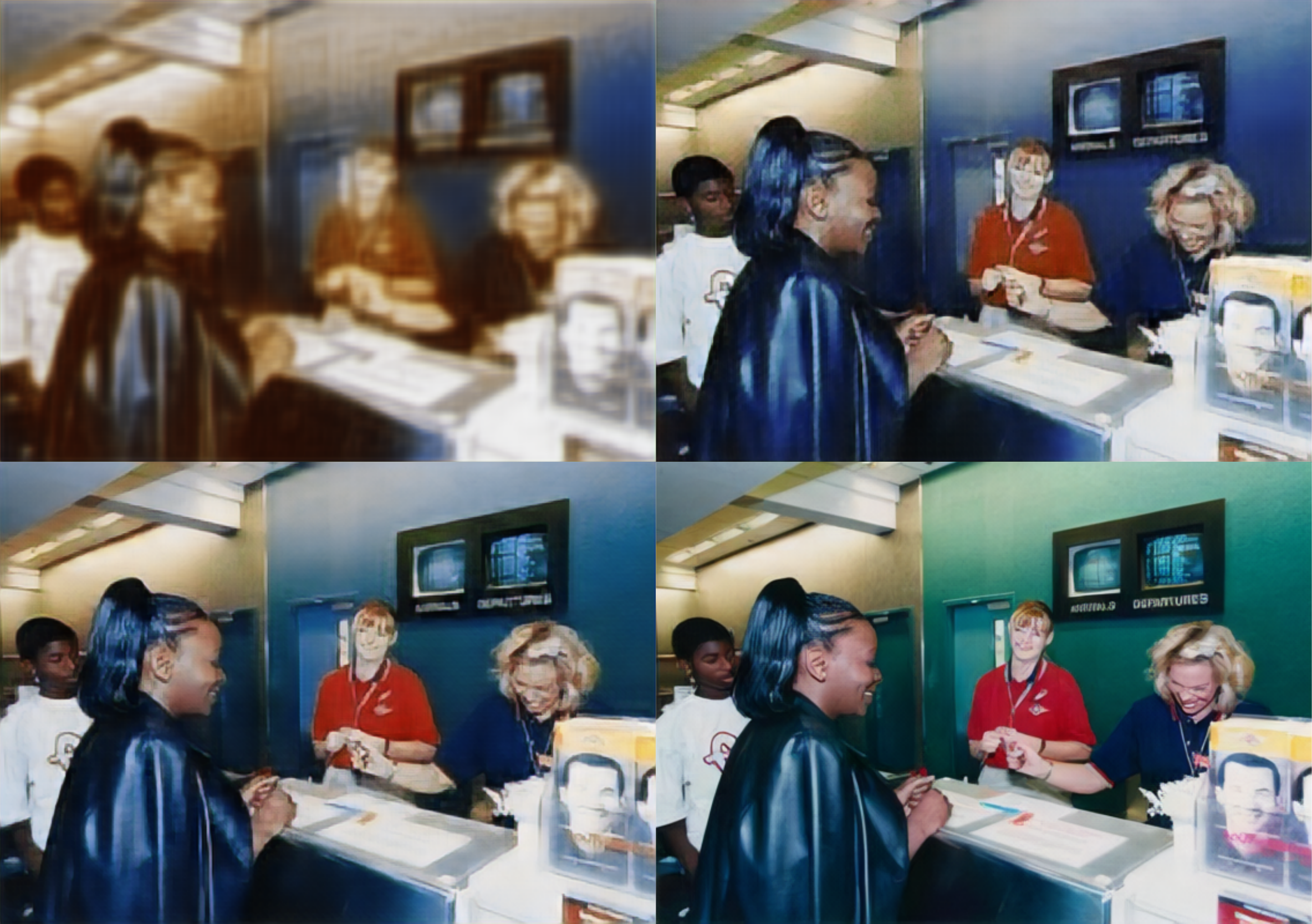}

  \caption{Estimated images of $\DSC$ for ADE20K dataset at $0.26~ \tBPP$ for varying number of training images. Left-top: $16$. Right-top: $56$. Left-bottom: $128$. Right-bottom: $8992$.}
   \label{fig:All_ADE20K_Original}
\end{figure}

In this section, we show $\DSC$ can also be quickly updated with low number of training images. Numerical and visual results are presented in Table \ref{tab:LowDatasets}, and Figs. \ref{fig:All01and05} and \ref{fig:All_ADE20K_Original}. Pre-trained RGAN for Cityscapes experiments is obtained from the ${\tPetr = 16}$ training at $0.24$ BPP seen in Table \ref{tab:VaryingCityscapes}. As detailed earlier, only BResNet and SumNet are trained when a pre-trained RGAN is used. Note that the disjoint $\DSC$ design achieves a higher PSNR than the joint $\DSC$ design, i.e., $40.07$ in Table \ref{tab:LowDatasets} vs. $39.49$ in Table \ref{tab:VaryingCityscapes} since RGAN output is higher in the former case, i.e., $\tPSNR(x, x') = 31.01$ vs. $27.76$. 

As seen in Figs. \ref{fig:All01and05} and \ref{fig:All_ADE20K_Original}, depending on the application, the number of training images can be kept at a significantly low value for the swift update of $\DSC$. Especially, $\DSC$ is very promising if the details are not that critical as seen in Fig. \ref{fig:All01and05}.

\begin{figure}[!t]
\centering
\includegraphics[scale=0.05] {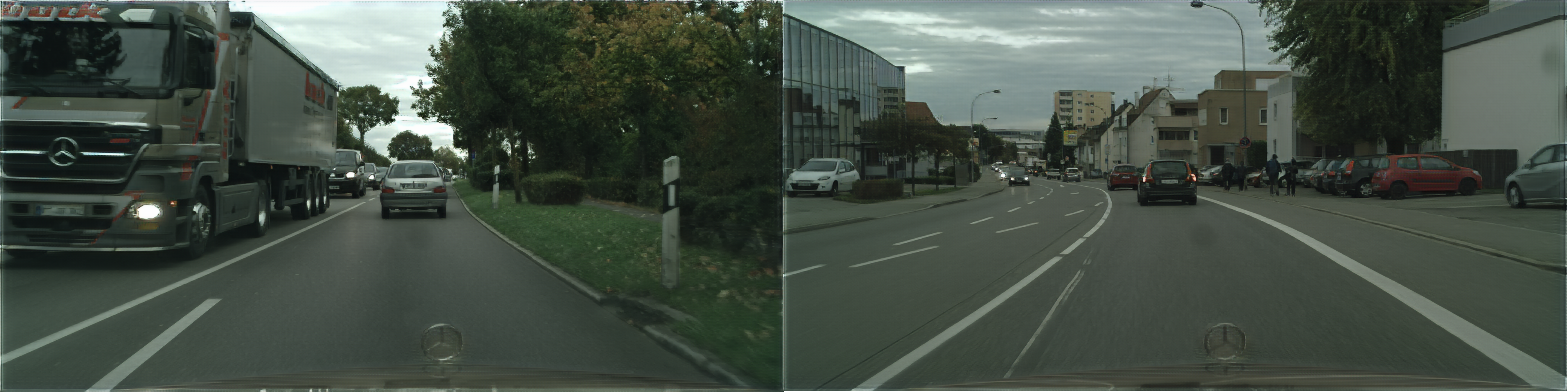}

  \caption{Synthesized images of $\DSC$ for Cityscapes dataset at $0.54~ \tBPP$.}
   \label{fig:All_Synth}
\end{figure}

In Fig. \ref{fig:All_Synth}, synthesized images $x^\prime$ for the Cityscapes dataset are shown. Depending on the application, only compressed $c$ and semantic label $s$ images can be transmitted if the estimation quality of synthesized image $x^\prime$ is sufficient for the application, e.g., satellite communications. As mentioned earlier, $c$ and $s$ are highly compressed images. Therefore, it is assumed they can be strongly protected and received error-free in this work. 

For Cityscapes, $c$ and $s$ file sizes are each about $18$ KB and the residual image $r$ is about $3$ MB. In other words, compared to $c$ and $s$, the bit rate of the $r$ contributes the most to the overall bit rate. The bit rate distribution among $r$, $c$, and $s$ for the Cityscapes dataset, as previously presented in Table \ref{tab:VaryingCityscapes}, is detailed in Table \ref{tab:BPPs}.

\begin{table}[!t]
  \begin{center}
    \caption{Results with low number of training images for Cityscapes (pre-trained RGAN is used) and ADE20K datasets at $0.54$ and $0.26$ BPPs, respectively.}
  \begin{tabular}{| c |c c c |}\hline
\textbf{Cityscapes} & $\boldsymbol\#$ \textbf{training images} & $\mathbf{PSNR(\boldsymbol{x,x^\prime})}$ & $\mathbf{PSNR(\boldsymbol{x},\hat{\boldsymbol{x}})}$   \\ \hline
 & $16$ & $31.01$ & $32.01$  \\ 
 & $88$ & $31.01$ & $36.8$  \\
 & $176$ & $31.01$ & $37.29$  \\
 & $2968$ & $31.01$ & $40.07$ \\ \hline \hline

\textbf{ADE20K} & $\boldsymbol\#$ \textbf{training images} & $\mathbf{PSNR(\boldsymbol{x,x^\prime})}$ & $\mathbf{PSNR(\boldsymbol{x},\hat{\boldsymbol{x}})}$   \\ \hline
              & $16$                           & $16.75$                         & $19.71$                        \\ 
& $56$ & $17.37$ & $24.72$ \\
& $128$ & $17.72$ & $25.53$ \\
& $8992$ & $22.03$ & $28.99$ \\ \hline
  \end{tabular}
      \label{tab:LowDatasets}
 \end{center}
\end{table}

\begin{table}[!t]
  \begin{center}
    \caption{The contributions of $r$, $c$, and $s$ to the total BPP for the Cityscapes dataset.}
  \begin{tabular}{| c | c | c |}\hline
$\boldsymbol r$ & $\boldsymbol c$ \textbf{ and }  $\boldsymbol s$ & \textbf{Total}  \\ \hline
 0.0625 & 0.0475  & 0.11  \\ \hline
0.125 & 0.045 & 0.17  \\  \hline
0.1875 & 0.0525 & 0.24  \\  \hline
0.5 & 0.04 & 0.54  \\  \hline
  \end{tabular}
      \label{tab:BPPs}
 \end{center}
\end{table}

\section{Conclusion}

In this work, we present a robust deep learning (DL) based semantic-empowered multi-layer image transmission scheme that is resilient to end-to-end (E2E) channel errors. With moderate changes to èxisting standards, the proposed deep semantics source channel coding ($\DSC$) solution can be viably implemented at the application layer (AL). We believe that flexibility is as important as achieving high estimation quality at high compression region. Thus, the proposed disjoint design of layered compressed image transmission offers adaptability across various scenarios under channel errors. Numerical results demonstrate the E2E channel error robustness of $\DSC$. Even when $\DSC$ is trained without channel errors, it experiences only an $11.4\%$ PSNR loss during the inference phase, and just a $3.3\%$ loss when trained with channel errors. Enhancing the performance of the proposed solution under ideal channel conditions is deferred to future work.

%
%
\bibliographystyle{splncs04}
\bibliography{SemanticComm}
\end{document}